\documentclass{IEEEtran4PSCC}

\ifCLASSINFOpdf
   \usepackage[pdftex]{graphicx}
\else
   \usepackage[dvips]{graphicx}
\fi

\usepackage[cmex10]{amsmath}

\usepackage[utf8]{inputenc}
\usepackage[margin=1in]{geometry}
\usepackage{enumitem}
\usepackage{amsmath,amsfonts,amssymb,mathtools}

\usepackage{graphicx,float}

\usepackage{listings}
\usepackage[ruled,vlined]{algorithm2e}
\usepackage{algorithmic}

\usepackage[colorlinks = true,
            linkcolor = blue,
            urlcolor  = blue,
            citecolor = blue,
            anchorcolor = blue]{hyperref}

\usepackage{tikz}
\usepackage{verbatim}
\usepackage{cite}

\newcommand{\exago}{ExaGO~}
\newcommand{\scopflow}{SCOPFLOW~}
\newcommand{\sopflow}{SOPFLOW~}
\newcommand{\pflow}{PFFLOW~}

\newcommand{\ipopt}{Ipopt~}
\newcommand{\tcopflow}{TCOPFLOW~}
\newcommand{\opflow}{OPFLOW~}
\newcommand{\hiop}{HiOp~}
\newcommand{\tao}{TAO~}
\newcommand{\raja}{\href{https://github.com/LLNL/RAJA}{RAJA \cite{beckingsale2019raja}}}
\newcommand{\umpire}{\href{https://github.com/LLNL/Umpire}{Umpire \cite{beckingsale2019umpire}}}

\makeatletter
\let\old@ps@headings\ps@headings
\let\old@ps@IEEEtitlepagestyle\ps@IEEEtitlepagestyle
\def\psccfooter#1{%
    \def\ps@headings{%
        \old@ps@headings%
        \def\@oddfoot{\strut\hfill#1\hfill\strut}%
        \def\@evenfoot{\strut\hfill#1\hfill\strut}%
    }%
    \def\ps@IEEEtitlepagestyle{%
        \old@ps@IEEEtitlepagestyle%
        \def\@oddfoot{\strut\hfill#1\hfill\strut}%
        \def\@evenfoot{\strut\hfill#1\hfill\strut}%
    }%
    \ps@headings%
}
\makeatother

\begin{document}

\title{Exascale Grid Optimization (ExaGO) toolkit: An open-source high-performance package for solving large-scale grid optimization problems
\thanks{This research was supported by the Exascale Computing Project (ECP), Project Number: 17-SC-20-SC, a collaborative effort of two DOE organizations—the Office of Science and the National Nuclear Security Administration—responsible for the planning and preparation of a capable exascale ecosystem—including software, applications, hardware, advanced system engineering, and early testbed platforms—to support the nation's exascale computing imperative.}
}

%% To specify the authors when (number of affiliations <= 2)
\author{
\IEEEauthorblockN{Shrirang Abhyankar \\ Slaven Pele\v{s} \\ Tamara Be\v{c}ejac \\ Jesse Holzer}
\IEEEauthorblockA{Electricity Infrasatructure and Buildings Division \\ Pacific Northwest National Laboratory (PNNL) \\
Richland, Washington, USA\\
\{shrirang.abhyankar, slaven.peles\}@pnnl.gov \\
\{tamara.becejac, jesse.holzer\}@pnnl.gov}
\and
\IEEEauthorblockN{Asher Mancinelli\\ Cameron Rutherford}
\IEEEauthorblockA{Research Computing \\ Pacific Northwest National Laboratory (PNNL) \\
Richland, WA, USA\\
\{asher.mancinelli, robert.rutherford\}@pnnl.gov}
}

% make the title area
\maketitle

\begin{abstract}
    This paper introduces the Exascale Grid Optimization (ExaGO) toolkit, a library for solving large-scale alternating current optimal power flow (ACOPF) problems including stochastic effects, security constraints and multi-period constraints. ExaGO can run on parallel distributed memory platforms, including massively parallel hardware accelerators such as graphical processing units (GPUs). We present the details of the ExaGO library including its architecture, formulations, modeling details, and its  performance for several optimization applications.
\end{abstract}

\section{Introduction}
The electric power grid remains vulnerable to disruption from extreme events including wildfires, severe storms, and cyber-attacks. Variable generation resources, load volatility and system being operated closer to its limits, also present operational challenges to grid stability. To mitigate disruptions before they snowball, grid planners and operators must be able to see these events coming and understand their potential impacts on grid reliability. Planners and operators rely on optimization and real-time tools to efficiently plan and operate the grid in a secure, cost-effective manner. The foundational bedrock for this analysis has been the optimal power flow (OPF) analysis \cite{Carpentier1979,Tinney1968,Cain2012}, which to date remains a challenging problem due to its non-convex nature. 

\begin{figure}[!h]
\centering
\includegraphics[width=\columnwidth]{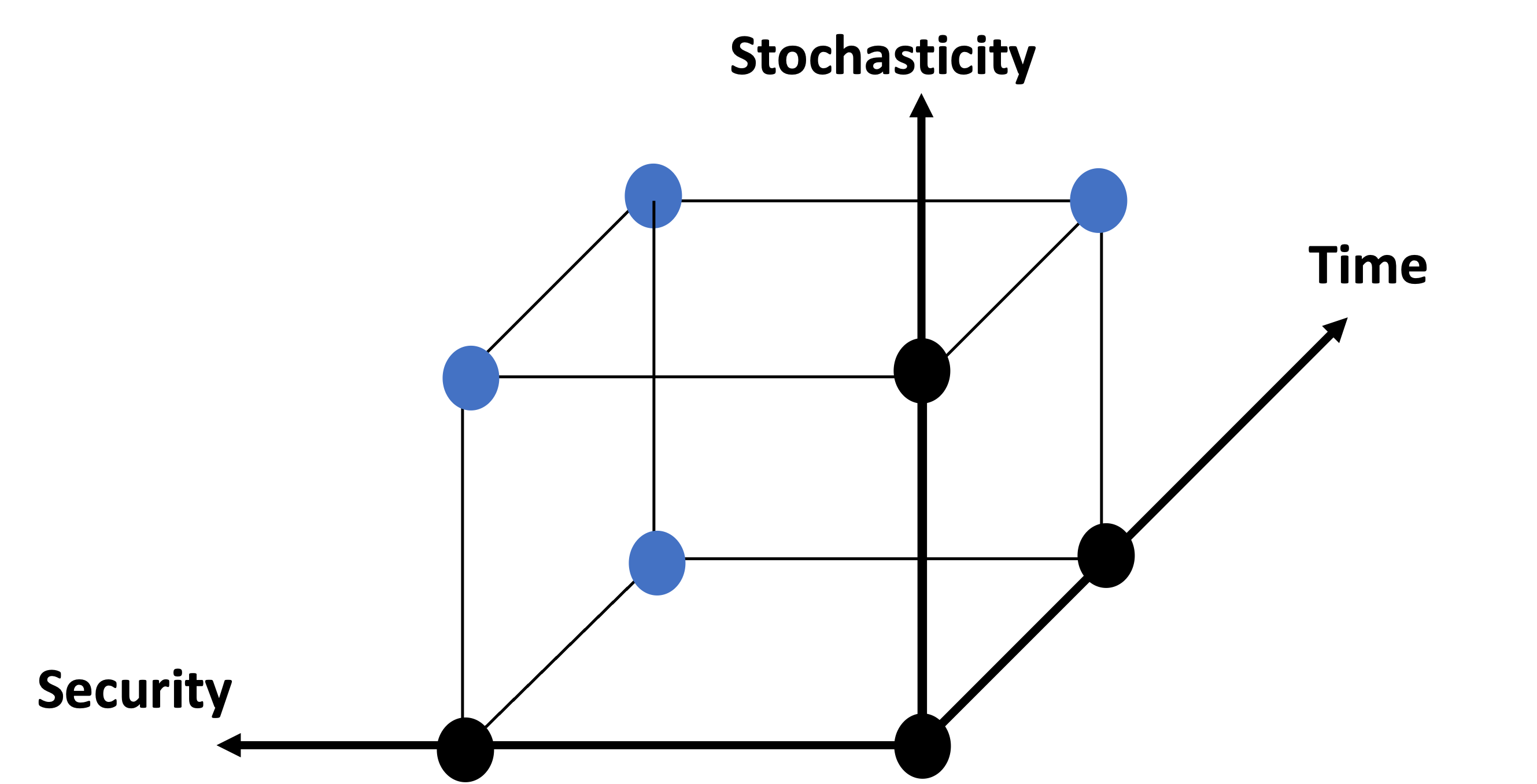}
\caption{The three uncertainty axes for the optimal power flow problem -- stochasticity, security, and time. The \textcolor{black}{black} nodes
represent optimal power flow problems along a single dimnesion, for e.g., stochastic optimal power flow, while the \textcolor{blue}{blue} nodes require a combination of dimensions, for e.g., stochastic security-constrained optimal power flow. Tools addressing OPF problems along two or more dimensions at scale are needed for the future low-inertia power systems with deep penetration of renewable energy sources}
\label{fig_sim}
\end{figure}

Over the years, the complexity of the optimal power flow grew as the grid's complexity increased and stringent reliability policies being mandated. Some examples include the incorporation of stochasticity\cite{Yong2000,Luo2018}, $N-1$ security constraints\cite{Capitanescu2008,Capitanescu2011}, multi-period aspects \cite{Schanen2018,Fuchs2017}, and a number of modeling complexities \cite{Farivar2013, Lubin2019}. Figure \ref{fig_sim} illustrates these three dimensions of the optimal power flow complexity.

To continue to ensure secure system operation, new tools that account for higher-level uncertainties, and are able to produce more accurate security assessment are required. Besides uncertainties, short-term decision will require additional attention, due to the more frequency ramping events caused by non-conventional generation, thus modeling that looks ahead a few steps will be desirable. Existing tools today either only address problems along a certain dimension, for e.g., security-constrained OPF only. Moreover, the scalability of such tools is limited only to a handful of "what-if condition" cases. 

In this paper, we introduce a novel open-source tool, the Exascale Grid Optimization Toolkit (ExaGO)\cite{exago_repo,exago_manual}, for solving Alternate Current OPF (ACOPF)-based optimization problems. The ExaGO toolkit can solve stochastic, security-constrained, and multiperiod ACOPF problems on high-performance with CPUs and GPUs. It contains of an array of applications to solve problems along a specific dimension (for e.g. Security-constrained ACOPF) or a combination of dimensions (for e.g., stochastic security-constrained ACOPF). 

ExaGO supports solving grid optimization problems expressed in general form given in (1)-(6). Here, the subscripts $s$, $c$, and $t$ represent the dimensions of stochasticity, security, and time, respectively.
\begin{align}
\centering
\text{min}&~\sum_{s \in \mathcal{S}}\pi_s\sum_{c \in \mathcal{C}}\sum_{t \in \mathcal{T}}f(x_{s,c,t})&  \label{eq:sctopflow_start}\\
&\text{s.t.}& \nonumber \\
&~g(x_{s,c,t}) = 0                                        & \\
&~h(x_{s,c,t}) \le 0                                      & \\
x^- & \le x_{s,c,t} \le x^+                               & \\
-\delta_t{x} & \le x_{s,c,t} - x_{s,c,t-\Delta{t}} \le \delta_t{x},~t \neq 0& \label{eq:sctopflow_time_coupling}\\
-\delta_c{x} & \le x_{s,c,0} - x_{s,0,0} \le \delta_c{x},~c \neq 0,&
\label{eq:sctopflow_contingency_coupling} \\
-\delta_s{x} & \le x_{s,0,0} - x_{0,0,0} \le \delta_s{x},~s \neq 0&
\label{eq:sctopflow_end}
\end{align}
where the subscripts $s \in \mathcal{S}$, $c \in \mathcal{C}$, and $t \in \mathcal{T}$ represent the dimensions of stochasticty, contingencies, and time, respectively, and $\pi_s$ is the probabiity or the weight of scenario $s$. The subscript $0$
denotes base-case for scenario or base-case for a contingency or the first time-period. The last three equations express the coupling between the scenarios, time, and contingencies, respectively.

ExaGO is written in C/C++ and uses high-performance computing (HPC) for scaling to solve large-scale problems. It can also be used with NViDIA and AMD graphical processing units (GPUs). Table \ref{tab:exago_comp} highlights the comparison of ExaGO with other open-source grid optimization tools - MOST \cite{most2020}, MATPOWER \cite{Zimmerman2011}, Pandapower \cite{pandapower}, PowerModels.jl \cite{powermodels}.

\begin{table}[h!]
    \caption{Comparison between ExaGO and other grid optimization tools}
    \centering
    \begin{tabular}{|p{0.2\columnwidth}|p{0.1\columnwidth}|p{0.1\columnwidth}|p{0.1\columnwidth}|p{0.1\columnwidth}|p{0.1\columnwidth}|}
        \hline
        & \textbf{ExaGO} &  \textbf{\cite{most2020}} & \textbf{\cite{Zimmerman2011}} & \textbf{\cite{pandapower}} & \textbf{\cite{powermodels}} \\ \hline
        Formulation & AC & DC & AC/DC & AC/DC & AC \\ \hline
        Language & C/C++ & Matlab & Matlab & Python & Julia \\ \hline
        Contingencies & Y & Y & N & N & N \\ \hline
        Multi-period & Y & Y & N & N & N \\ \hline
        Stochasticity & Y & Y & N & N & N \\ \hline
        Parallel computing & Y & N & N & N & N \\ \hline
        GPU & Y & N & N & N & N \\ \hline
    \end{tabular}
    \label{tab:exago_comp}
\end{table}

\section{ExaGO architecture}\label{sec:exago_arch}
 ExaGO is written in C/C++ and makes heavy use of the functionality provided by the PETSc\cite{petsc-user-ref} library. All ExaGO applications use a nonlinear formulation based on full AC optimal power flow. The different applications available with \exago are listed in Table \ref{tab:exago_apps}

\begin{table}[!htbp]
    \centering
  \caption{\exago applications}
  \begin{tabular}{|p{0.3\columnwidth}|p{0.5\columnwidth}|}
    \hline
    \textbf{Application Name} & \textbf{Description} \\
    \hline
    \opflow & AC optimal power flow \\ \hline
    \tcopflow & Multi-period AC optimal power flow \\ \hline
    \scopflow & Single/multi-period security-constrained AC optimal power flow \\ \hline
    \sopflow & Single/multi-period no/multi-contingency stochastic AC optimal power flow  \\
    \hline
    \pflow & AC power flow  \\ \hline
  \end{tabular}
  \label{tab:exago_apps}
\end{table}

\begin{figure}[h!]
\centering
\includegraphics[width=\columnwidth]{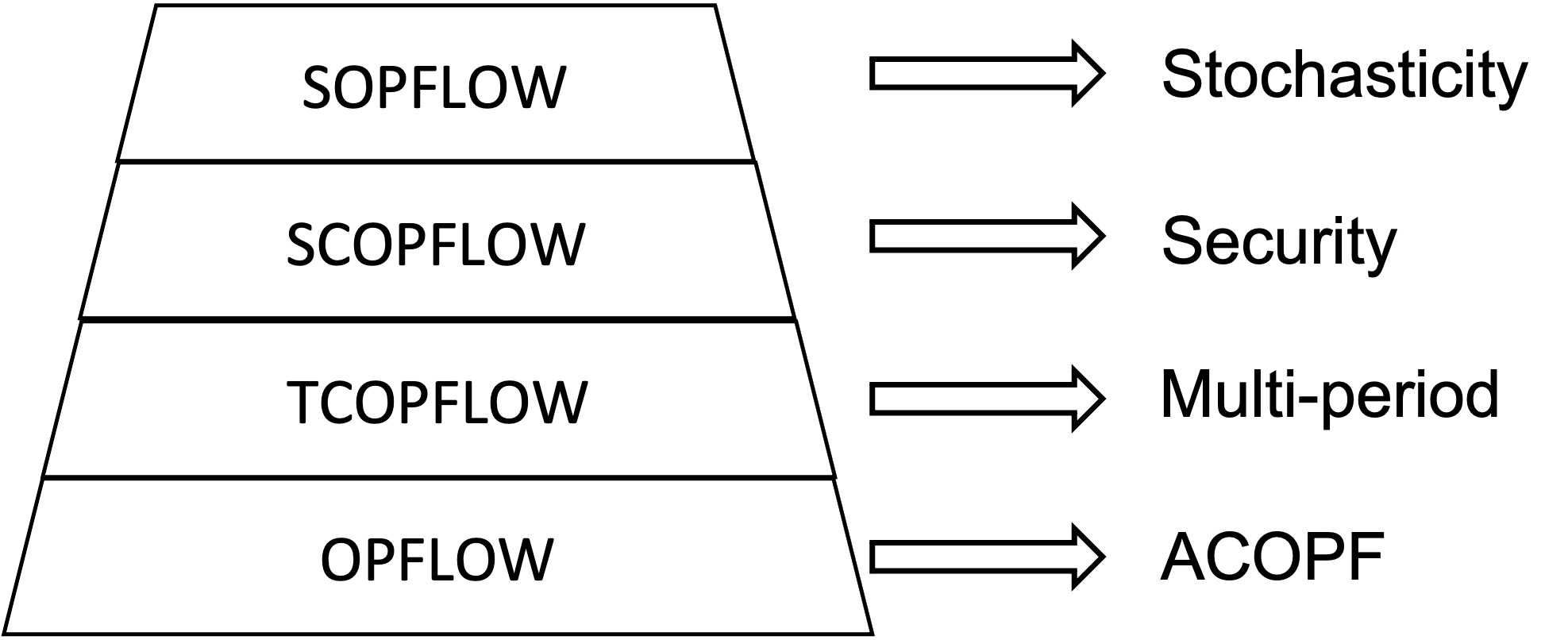}
\caption{ExaGO application hierarchy}
\label{fig:app_hierarchy}
\end{figure}

The hierarchy of the applications is shown in Figure \ref{fig:app_hierarchy} to show how the applications are layered.

\begin{figure}[!h]
\centering
\includegraphics[width=\columnwidth]{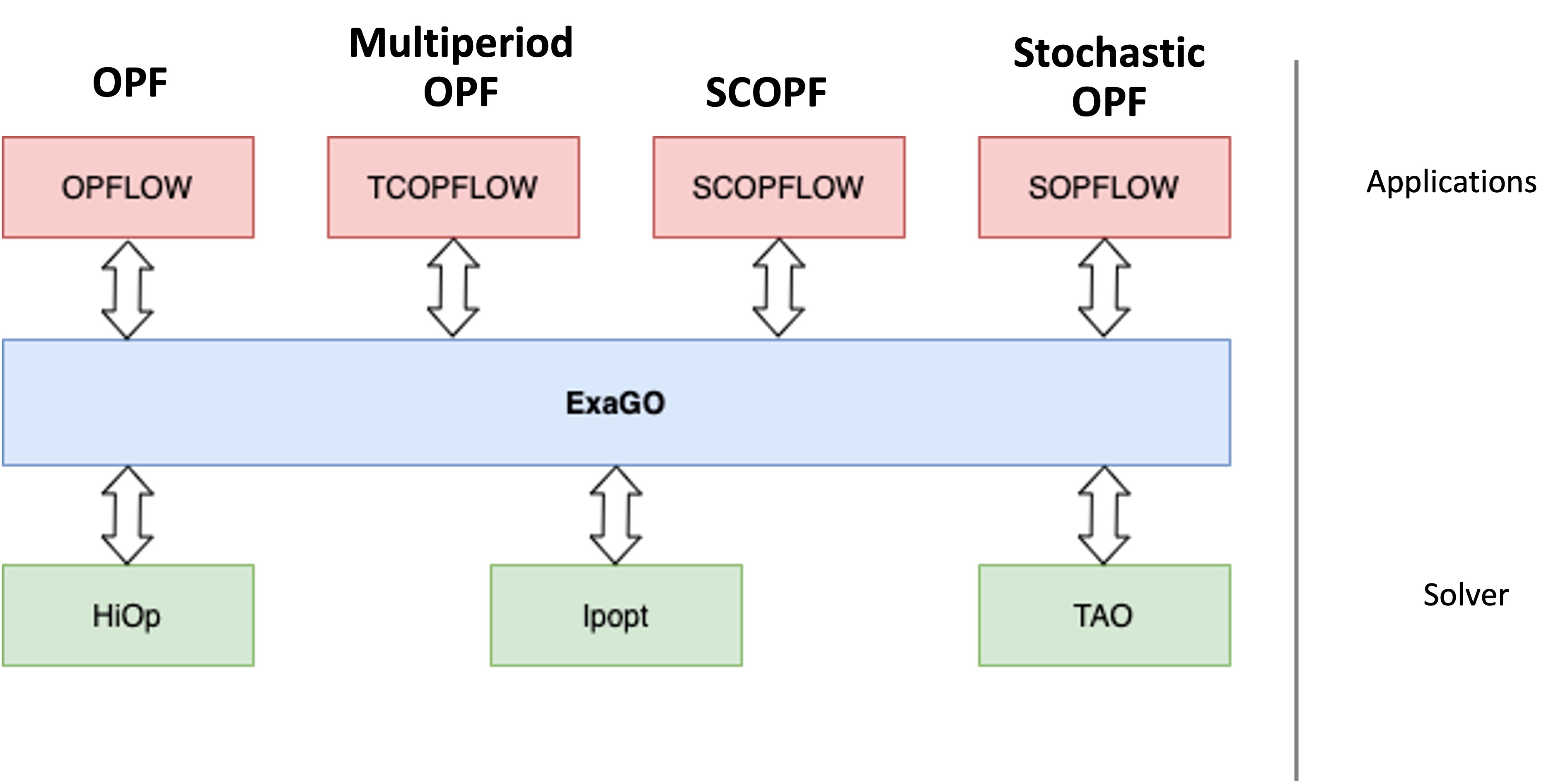}
\caption{ExaGO applications and solvers}
\label{fig:apps_solvers}
\end{figure}

ExaGO is interfaced with three optimization solvers -- \ipopt, \hiop, and \tao -- for solving the underlying optimization problems. Figure \ref{fig:apps_solvers} shows the solvers available for the different applications.

One unique computational feature of ExaGO is the ability to perform calculations (model evaluation and optimization solves) on the GPU. For this, it uses \hiop, \raja, and \umpire~libraries. All the calculations are done directly on the GPU, i.e., there is minimal exchange of data between CPU and GPU during the optimization iterations. The reader is refered \cite{exago_gpu} for GPU implementation details.

In the next sections, we describe the different ExaGO applications.

\section{OPFLOW: AC optimal power (ACOPF)}
OPFLOW solves the full AC optimal power flow problem and provides various flexible features that can be toggled via run-time options. It has interfaces to different optimization solvers and includes different representations of the underlying equations (power-balance-polar and power-balance-cartesian) that can be used. By selecting the appropriate solver, OPFLOW can be executed on CPUs (in serial or parallel) or on GPUs.

\begin{figure}[h!]
\centering
  {\includegraphics[width=\columnwidth]{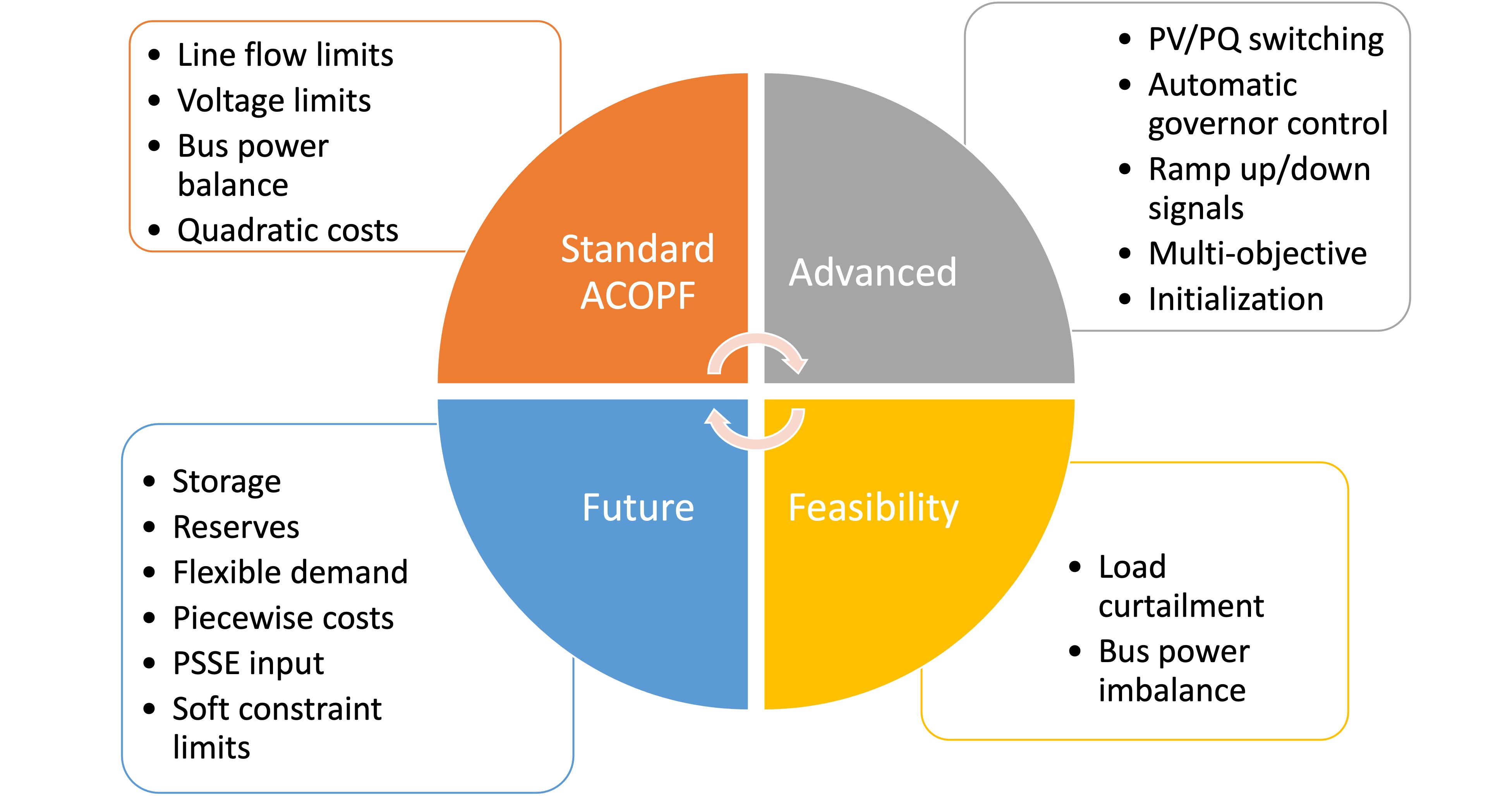}}
  \caption{ExaGO formulation includes standard ACOPF formulation plus feasabiity (load loss, power imbalance), and advanced features (PV/PQ switching, AGC)}
  \label{fig:acopf_modeling}
\end{figure}

In compact form, the set of equations describing ACOPF is given by (\ref{eq:opfobj})-(\ref{eq:opfbounds}).
\begin{align}
\underset{x}{\text{min}}&~ f(x)& \label{eq:opfobj}\\
&\text{s.t.}& \nonumber \\
&g(x) = 0& \label{eq:opfeq}\\
h^{-}& \le h(x) \le h^+& \label{eq:opfineq}\\
x^{-}& \le x \le x^{+}& \label{eq:opfbounds}
\end{align}
The full formulation of an AC optimal power flow is digressed in this paper due to space limitations. The reader is referred to \cite{exago_manual,Zimmerman2011,ONeill2012} for comprehensive details on the AC optimal power flow formulation. Figure \ref{fig:acopf_modeling} briefly describes the modeling details of OPFLOW. 

%\todo{quick description of modeling features: load loss, power imbalance, initialization types, inputs and outputs.} The reader is refered to the ExaGO manaul \cite{exago_manual} for comprehensive details of the OPFLOW formulation. 

\begin{figure}[h!]
\centering
  {\includegraphics[width=\columnwidth]{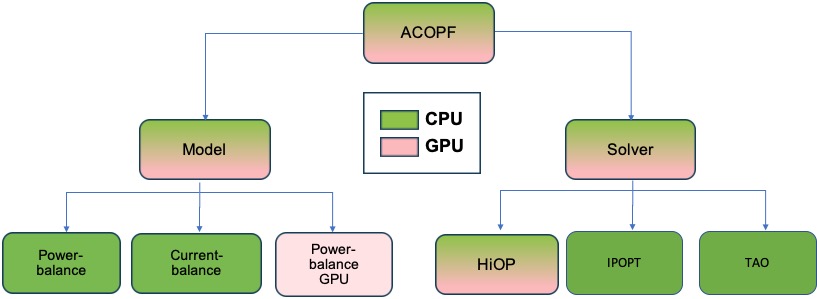}}
  \caption{Architecture of the ACOPF application in ExaGO}
  \label{fig:acopf_arch}
\end{figure}

The layout of the ACOPF application in ExaGO is shown in Fig. \ref{fig:acopf_arch}. A clear separation of model (physics) and the solver is implemented to allow switching between different models/solvers (through run-time options). OPFLOW can be used with three different solvers: \ipopt \cite{ipopt}, \hiop \cite{hiop-manual,hiop1}, and \tao \cite{petsc-user-ref}.

 Different colors in Fig. \ref{fig:acopf_arch} denote models/solvers compatible with CPU and GPU, respectively. The advantage of such a separation is that hardware-specific code for a model or a solver is isolated to it and does not spill over other portions of the code base and can be easily replaced.

\section{TCOPFLOW: Multi-period OPF}\label{sec:tcopflow}

TCOPFLOW solves a full AC multi-period optimal power flow problem with the objective of minimizing the total cost over the given time horizon while adhering to constraints for each period and between consecutive time-periods (ramping constraints). 

The multi-period optimal power flow problem is a series of optimal power flow problems coupled via temporal constraints. The generator real power deviation ($p_{jt}^{\text{g}} - p_{jt-\Delta{t}}^{\text{g}}$) constrained within the ramp limits form the temporal constraints. 

An illustration of the temporal constraints is shown in Fig. \ref{fig:tcopflow} with four time steps. Each time-step $t$ is coupled with its preceding time $t-\Delta{t}$, where $\Delta{t}$ is the time-step where the objective is to find a least cost dispatch for the given time horizon.
\definecolor{lavander}{cmyk}{0,0.48,0,0}
\definecolor{violet}{cmyk}{0.79,0.88,0,0}
\definecolor{burntorange}{cmyk}{0,0.52,1,0}

\def\lav{lavander!90}
\def\oran{orange!30}

\tikzstyle{time}=[draw,circle,violet,bottom color=\lav,
                  top color= white, text=violet,minimum width=20pt]
\tikzstyle{base}=[draw,circle,burntorange, left color=\oran,
                       text=violet,minimum width=20pt]

\begin{figure}[h!]
\centering
\begin{tikzpicture}[auto, thick]
  \node[time,label=below:$t_0$] (t0) at (0,0) {};
  \node[time,label=below:$t_1$] (t1) at (2,0) {};
  \node[time,label=below:$t_2$] (t2) at (4,0) {};
  \node[time,label=below:$t_3$] (t3) at (6,0) {};
  
  \path (t0) edge (t1);
  \path (t1) edge (t2);
  \path (t2) edge (t3);

\end{tikzpicture}
\caption{Multi-period optimal power flow example with four time-steps. The lines connecting the different time-periods denote the coupling between them.}
\label{fig:tcopflow}
\end{figure}
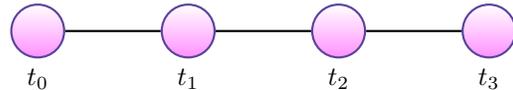

In general form, the equations for multi-period optimal power flow are given by (\ref{eq:tcopflow_start})-(\ref{eq:tcopflow_end}). TCOPFLOW solves to minimize the total generation cost $\sum_{t \in \mathcal{T}}f(x_t)$ over the time horizon $T$. At each time-step, the equality constraints ($g(x_t)$), inequality $h(x_t)$, and the lower/upper limit ($x^-$,$x^+$) constraints need to be satisfied. Equation (\ref{eq:tcopflow_end}) represents the coupling between the consecutive time-steps. The most common form of coupling are ramping constraints that limit the deviation of the real power generation at time $t$ from its preceding time-step $t-\Delta{t}$ to within its ramping capability $\delta_t{x}$.

\begin{align}
\centering
\text{min}&~\sum_{t \in \mathcal{T}} f(x_t)  \label{eq:tcopflow_start}\\
&\text{s.t.}& \nonumber \\
&~g(x_t) = 0  \\
&~h(x_t) \le 0 \\
x^- & \le x_t \le x^+ \\
-\delta_t{x} & \le x_t - x_{t-\Delta{t}} \le \delta_t{x},&t \neq 0 &
\label{eq:tcopflow_end}
\end{align}

\begin{figure}[h!]
\centering
  {\includegraphics[width=\columnwidth]{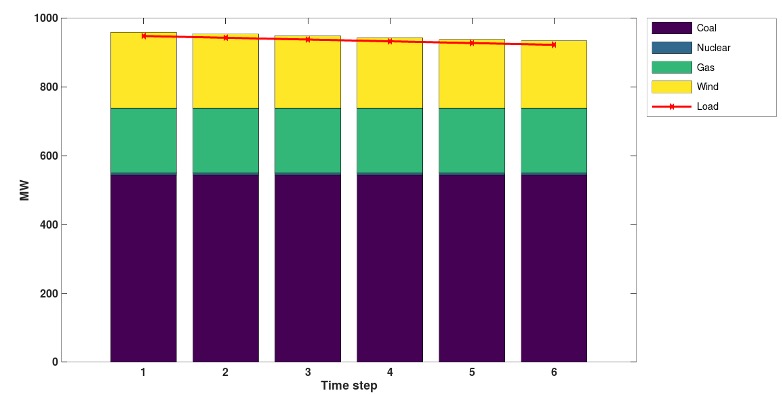}}
  \caption{Generation dispatch for ACTIVSg200 \cite{activsg1} bus system from TCOPLOW for a 30-min horizon with 5-minute intervals}
  \label{fig:tcopflow_gen}
\end{figure}

\tcopflow currently supports solving the multi-period OPF problem using \ipopt on single processor only. Figure \ref{fig:tcopflow_gen} shows the results from \tcopflow for a 30-min. horizon with 5 minute intervals for the ACTIVSg200 \cite{activsg1} bus system.

Load and/or wind generation profiles can be set for the multiperiod probem with \tcopflow. The output of the base-case and the contingency subproblems is saved to MATPOWER format files (one file per subproblem).

%\begin{itemize}
%    \item \textbf{Network file:} The network file describing the network details. Only \matpower format files are currently supported.
%    \item \textbf{Load data:} One file for load real power and one for reactive power. The files need to be in CSV format. An example of the format for the 9-bus case is \href{https://gitlab.pnnl.gov/exasgd/frameworks/exago/-/tree/master/datafiles/case9}{here}.
%    \item \textbf{Wind generation:} The wind generation time-series described in CSV format. See an example of the format \href{https://gitlab.pnnl.gov/exasgd/frameworks/exago/-/tree/master/datafiles/case9}{here}.
%\end{itemize}
%If the load data and/or wind generation profiles are not set then a flat profile is assumed, i.e., the load and wind generation for all hours is constant.

%The \tcopflow output is saved to a directory named \emph{tcopflowout}. This directory contains $N_t$ files, one for each time-step, in \matpower data file format.

\section{SCOPFLOW: Single/Multi-period Security-constrained ACOPF}\label{sec:scopflow}
\scopflow solves a single/multi-period contingency-constrained optimal power flow problem. The problem is set up as a two-stage optimization problem where the first-stage (base-case $c_0$) represents the normal operation of the grid and the second-stage comprises $c \in \mathcal{C}$ contingency cases. Each contingency case can be single or multi-period. \scopflow operates in two modes: (a) \emph{preventive} - only reference generators pick up the deficit/excess power for the contingency, or (b) \emph{corrective} - the generators are allowed to ramp up/dowm constrained by their ramping limits. In the next subsections, we described the single-period and multi-period formulations for \scopflow.

\subsection{Single-period}

The contingency-constrained optimal power flow (popularly termed as security-constrained optimal power flow (SCOPF) in power system parlance) attempts to find a least cost dispatch for the base case (or no contingency) while ensuring that if any of contingencies do occur then the system will be secure. This is illustrated in Fig. \ref{fig:scopflow} for a SCOPF with a base-case $c_0$ and three contingencies.

\definecolor{lavander}{cmyk}{0,0.48,0,0}
\definecolor{violet}{cmyk}{0.79,0.88,0,0}
\definecolor{burntorange}{cmyk}{0,0.52,1,0}

\def\lav{lavander!90}
\def\oran{orange!30}

\tikzstyle{contingency}=[draw,circle,violet,bottom color=\lav,
                  top color= white, text=violet,minimum width=20pt]
\tikzstyle{base}=[draw,circle,burntorange, left color=\oran,
                       text=violet,minimum width=20pt]
                       
\tikzstyle{cedge}=[color=red]

\begin{figure}[h!]
\centering
\begin{tikzpicture}[auto, thick]
  % Place base case
  \node[base,label=left:$c_0$] (base) at (0,0) {};
  
  \node[contingency,label=right:$c_1$] (c1) at (2,1) {};
  \node[contingency,label=right:$c_2$] (c2) at (2,0) {};
  \node[contingency,label=right:$c_3$] (c3) at (2,-1) {};
  
  \path[cedge] (base) edge (c1);
  \path[cedge] (base) edge (c2);
  \path[cedge] (base) edge (c3);

%  \foreach \place/\name in {{(0,-1)/a}, {(2,0)/b}, {(2,2)/c}, {(0,2)/d},
%           {(-2,0)/e}}
%    \node[superpeers] (\name) at \place {a};
%  \foreach \source/\dest in {a/b, a/c, a/d, b/c, c/d,a/e,d/e}
%    \path (\source) edge (\dest);
   %
   % Place normal peers
%  \foreach \pos/\i in {above left of/1, left of/2, below left of/3}
%    \node[peers, \pos = e] (e\i) {};
%   \foreach \speer/\peer in {e/e1,e/e2,e/e3}
%    \path (\speer) edge (\peer);
   %
%   \foreach \pos/\i in {above right of/1, right of/2, below right of/3}
%    \node[peers, \pos =b ] (b\i) {};
%   \foreach \speer/\peer in {b/b1,b/b2,b/b3}
%   \path (\speer) edge (\peer);
   %
%   \node[peers, above of=d] (d1){};
%   \path (d) edge (d1);
   %
%   \foreach \pos/\i in {below left of/1, below of/2}
%   \node[peers, \pos =a ] (a\i) {};
%   \foreach \speer/\peer in {a/a1,a/a2}
%   \path (\speer) edge (\peer);

\end{tikzpicture}
\caption{Contingency constrained optimal power flow example with three contingencies. $c_0$ represents the base case (or no contingency case). $c_1$, $c_2$, $c_3$ are the three contingency cases. Each of the contingency states is coupled with the base-case through ramping constraints (denoted by \textcolor{red}{red} lines)}
\label{fig:scopflow}
\end{figure}
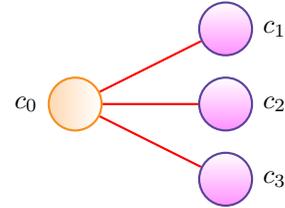

In general form, the equations for contingency-constrained optimal power flow are given by
(\ref{eq:scopflow_start}) -- (\ref{eq:scopflow_end}). This is a two-stage stochastic optimization problem where the first stage is the base case $c_0$ and each of the contingency states are second-stage subproblems.

SCOPFLOW aims to minimize the objective $\sum_{c \in \mathcal{C}}f(x_c)$, while adhering to the equality $g(x_c)$, inequality $h(x_c)$, and the lower/upper bound ($x^-$,$x^+$) constraints. For notational ease we include the base-case in set $\mathcal{C}$, i.e., $\mathcal{C} \equiv \mathcal{C} \cup c_0$. Equation (\ref{eq:scopflow_end}) represents the coupling between the base-case and each of the contingency states $c_i$. Equation (\ref{eq:scopflow_end}) is the most typical form of coupling that limits the deviation of the contingency variables $x_c$ from the base $x_0$ to within $\delta_c{x}$. An example of this constraint could be the allowed real power output deviation for the generators constrained by their ramp limit.

\begin{align}
\centering
\text{min}&\sum_{c \in \mathcal{C}}f(x_c)  \label{eq:scopflow_start}\\
&\text{s.t.} \nonumber \\
&~g(x_c) = 0,                              \\
&~h(x_c) \le 0,                            \\
x^- & \le x_c \le x^+,                     \\
-\delta_c{x} & \le x_c - x_0 \le \delta_c{x},~ c\neq 0
\label{eq:scopflow_end}
\end{align}
where, $\mathcal{C}$ represents the set of contingencies, including the base-case denoted by subscript 0. The list of contingencies is provided through a PTI or native format file. Each contingency can be single outage (e.g. single generator outage) or multiple (e.g. simultaneous generator and a line outage).

\begin{figure}[h!]
\centering
  {\includegraphics[width=\columnwidth]{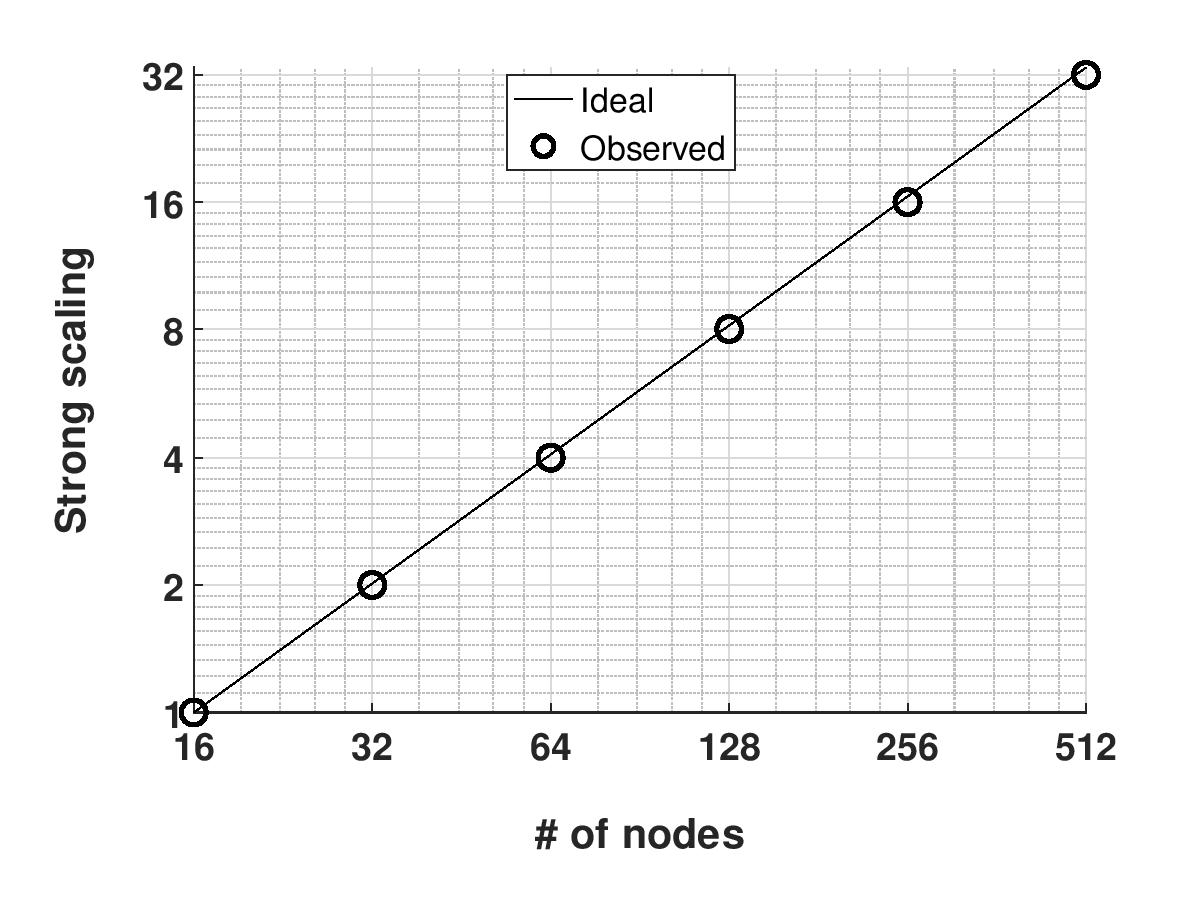}}
  \caption{Scalability of SCOPFLOW on 512 nodes (3096 ranks) on Summit cluster \cite{summit}. The embarrasingly parallel solver EMPAR was used for solving SCOPFLOW on the ACTIVSg2000 \cite{activsg1} system with 3095 contingencies.}
  \label{fig:scopflow_empar_scaling}
\end{figure}

Three different solvers are available to solve single-period \scopflow:
\begin{itemize}
    \item With the \ipopt solver, \scopflow constructs a large monolithic problem consisting of the base case and the contingencies. This solver is only supported in serial.
    \item The \hiop solver uses a primal decomposition approach to parallelize SCACOPF analysis where each contingency problem (and base case) are solved independently. \hiop uses a primal decomposition approach with second order corrections \cite{hiop-repo} to solve the \scopflow problem. This solver is supported in serial and parallel, and can also use GPUs for solving the contingency subproblems. Figure \ref{fig:scopflow_hiop_scaling} shows strong scaling of SCOPFLOW with HiOP primal decomposition algorithm on the ACTIVSg200 \cite{activsg1} test system with 100 contingencies.
\begin{figure}[h!]
\centering
  {\includegraphics[width=\columnwidth, trim = 0cm 0cm 0cm 13cm,clip]{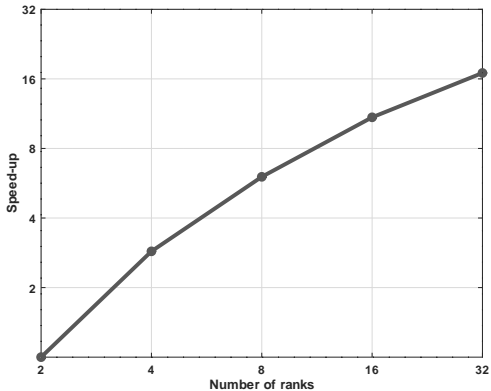}}
  \caption{Scalability of SCOPFLOW using primal decomposition approach. The HIOP solver was used for solving SCOPFLOW on the ACTIVSg200 system with 100 contingencies.}
  \label{fig:scopflow_hiop_scaling}
\end{figure}
    \item The EMPAR solver is an \emph{embarrasingly parallel} version of \scopflow that solves the base case and contingency subproblems independently in parallel. In other words, it relaxes the coupling constraint between the base-case and contingency subproblems and solves them separately. Figure \ref{fig:scopflow_empar_scaling} shows the scalability of the embarrasingly parallel solver for a large-scale optimization run.
\end{itemize}

\subsection{Multiperiod}

In the multi-period version of \scopflow, each contingency comprised of multiple time-periods. The multiple periods have variables and constraints as described in section \ref{sec:tcopflow}, i.e., each multiperiod problem uses TCOPFLOW application internally. An example of multi-contingency multi-period optimal power flow is illustrated in Fig. \ref{fig:ctopflow} with two contingencies $c_0$ and $c_1$. Here, $c_0$ is the case with no contingencies, i.e., the base-case. In this example, each contingency is multi-period with four time-periods. Each time-step is coupled with its adjacent one through ramping constraints. Currently, we assume the contingency is incident at the first time-step, i.e. at $t_0$. This results in the coupling between the contingency cases $c \in \mathcal{C}$ and the base-case $c_0$ only at time-step $t_0$ as shown in Fig. \ref{fig:ctopflow}.

\definecolor{lavander}{cmyk}{0,0.48,0,0}
\definecolor{violet}{cmyk}{0.79,0.88,0,0}
\definecolor{burntorange}{cmyk}{0,0.52,1,0}

\def\lav{lavander!90}
\def\oran{orange!30}

\tikzstyle{time}=[draw,circle,violet,bottom color=\lav,
                  top color= white, text=violet,minimum width=20pt]
                  
\tikzstyle{base}=[draw,circle,burntorange, left color=\oran,
                       text=violet,minimum width=20pt]

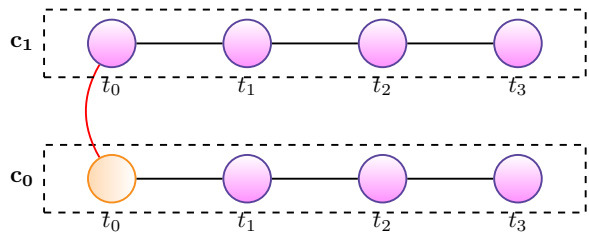
\begin{figure}[h!]
\centering
\scalebox{0.9}{
\begin{tikzpicture}[auto, thick]
  \node[base,label=below:$t_0$] (c0t0) at (0,0) {};
  \node[time,label=below:$t_1$] (c0t1) at (2,0) {};
  \node[time,label=below:$t_2$] (c0t2) at (4,0) {};
  \node[time,label=below:$t_3$] (c0t3) at (6,0) {};
  
  \path (c0t0) edge (c0t1);
  \path (c0t1) edge (c0t2);
  \path (c0t2) edge (c0t3);
  
  % rectangle from (-1.0,-0.5) to (7.0,0.5)
   \node[draw,rectangle,dashed,minimum width=8cm,minimum height=1cm,label=left:{$\mathbf{c_0}$}] (s0) at (3.0,0.0) {};
  
  \node[time,label=below:$t_0$] (c1t0) at (0,2) {};
  \node[time,label=below:$t_1$] (c1t1) at (2,2) {};
  \node[time,label=below:$t_2$] (c1t2) at (4,2) {};
  \node[time,label=below:$t_3$] (c1t3) at (6,2) {};
  
  \path (c1t0) edge (c1t1);
  \path (c1t1) edge (c1t2);
  \path (c1t2) edge (c1t3);
  
  \path (c0t0) edge [bend left,color=red] (c1t0);
  
   % rectangle from (-1.0,1.5) to (7.0,2.5)
   \node[draw,rectangle,dashed,minimum width=8cm,minimum height=1cm,label=left:{$\mathbf{c_1}$}] (s0) at (3.0,2.0) {};

%  \foreach \place/\name in {{(0,-1)/a}, {(2,0)/b}, {(2,2)/c}, {(0,2)/d},
%           {(-2,0)/e}}
%    \node[superpeers] (\name) at \place {a};
%  \foreach \source/\dest in {a/b, a/c, a/d, b/c, c/d,a/e,d/e}
%    \path (\source) edge (\dest);
   %
   % Place normal peers
%  \foreach \pos/\i in {above left of/1, left of/2, below left of/3}
%    \node[peers, \pos = e] (e\i) {};
%   \foreach \speer/\peer in {e/e1,e/e2,e/e3}
%    \path (\speer) edge (\peer);
   %
%   \foreach \pos/\i in {above right of/1, right of/2, below right of/3}
%    \node[peers, \pos =b ] (b\i) {};
%   \foreach \speer/\peer in {b/b1,b/b2,b/b3}
%   \path (\speer) edge (\peer);
   %
%   \node[peers, above of=d] (d1){};
%   \path (d) edge (d1);
   %
%   \foreach \pos/\i in {below left of/1, below of/2}
%   \node[peers, \pos =a ] (a\i) {};
%   \foreach \speer/\peer in {a/a1,a/a2}
%   \path (\speer) edge (\peer);

\end{tikzpicture}
}
\caption{Multi-period contingency constrained optimal power flow example with two contingencies $c_0$ and $c_1$, each with four time-periods $t_0$, $t_1$, $t_2$, $t_3$. State $c_0,t_0$ represent the base case (no contingency) case. We assume that any contingency is incident at the first time-step, i.e., at $t_0$. The contingency states $c_1,t_0$ is coupled with the no-contingency state $c_0,t_0$ at time $t_0$. The {\textcolor{red}{red}} line denotes the coupling between the contingency.}
\label{fig:ctopflow}
\end{figure}

The overall objective of this contingency-constrained multi-period optimal power flow is to find a secure dispatch for base-case $c_0$ while adhering to contingency and temporal constraints. The general formulation of this problem is given in Eqs. (\ref{eq:ctopflow_start}) -- (\ref{eq:ctopflow_end}).

\begin{align}
\centering
\text{min}&~\sum_{c \in \mathcal{C}}\sum_{t \in \mathcal{T}}f(x_{c,t}) \label{eq:ctopflow_start}\\
&\text{s.t.} \nonumber \\
&~g(x_{c,t}) = 0                                         \\
&~h(x_{c,t}) \le 0                                       \\
x^- & \le x_{c,t} \le x^+                                \\
-\delta_t{x} & \le x_{c,t} - x_{c,t-\Delta{t}} \le \delta_t{x},~t\neq 0 \label{eq:ctopflow_time_coupling}\\
-\delta_c{x} & \le x_{c,0} - x_{0,0} \le \delta_c{x},~c\neq 0 \\
\label{eq:ctopflow_end}
\end{align}
here, $\mathcal{C}$ and ${T}$ are the sets for contingencies (including the base-case) and time-steps, respectively.

The objective of the multi-period SCOPFLOW formulation is to reduce the total cost for the given horizon over the set of contingencies. Equation (\ref{eq:ctopflow_end}) represents the coupling between the base case $c_0$ and each contingency $c_i$ at time-step $t_0$. We use a simple box constraint $\delta_c{x}$ to restrict the  deviation of decision variables $x_{c,0}$ from the base-case $x_{0,0}$. The bound $\delta_c{x}$ could represent here, for example, the allowable reserve for each generator. All the modeling details of TCOPFLOW \ref{sec:tcopflow} can be used for this multi-period setup. The multi-period SCOPFLOW currently supports solution using \ipopt only. It can also do an embarrasingly parallel solution where the contingency coupling constraints are relaxed and each multi-period problem is distributed and solved independently on a rank.

\section{SOPFLOW: Single/multiperiod no/multi-contingency Stochastic ACOPF}\label{sec:sopflow}

\sopflow solves a stochastic optimal power flow problem where the stochasticity is described through scenarios of wind and/or load forecast deviations. The problem is set up as a two-stage optimization problem where the first-stage (base-case) represents the normal operation of the grid (or the most likely forecast) and the second-stage comprises $N_s$ scenarios of forecast deviation. Each scenario can have multiple contingencies and each contingency can be multi-period. Like \scopflow, \sopflow also operates in either a preventive or corrective mode of generation redispatch. 

In the next subsections, we describe the different variations of \sopflow.

\subsection{Single period, no contingencies}
This problem has a structure as illustrated in Fig. \ref{fig:sopflow}. 

\definecolor{lavander}{cmyk}{0,0.48,0,0}
\definecolor{violet}{cmyk}{0.79,0.88,0,0}
\definecolor{burntorange}{cmyk}{0,0.52,1,0}

\def\lav{lavander!90}
\def\oran{orange!30}

\tikzstyle{contingency}=[draw,circle,violet,bottom color=\lav,
                  top color= white, text=violet,minimum width=20pt]
\tikzstyle{base}=[draw,circle,burntorange, left color=\oran,
                       text=violet,minimum width=20pt]
                       
\tikzstyle{cedge}=[color=red]

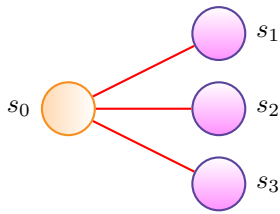
\begin{figure}[h!]
\centering
\begin{tikzpicture}[auto, thick]
  % Place base case
  \node[base,label=left:$s_0$] (base) at (0,0) {};
  
  \node[contingency,label=right:$s_1$] (s1) at (2,1) {};
  \node[contingency,label=right:$s_2$] (s2) at (2,0) {};
  \node[contingency,label=right:$s_3$] (s3) at (2,-1) {};
  
  \path[cedge] (base) edge (s1);
  \path[cedge] (base) edge (s2);
  \path[cedge] (base) edge (s3);
  
  \end{tikzpicture}
\caption{Stochastic optimal power flow example with three four. $s_0$ represents the base case (or the scenario with the highest probability). $s_1$, $s_2$, $s_3$ are the other three scenario forecasts. Each of the scenarios is coupled with the base-case scenario through ramping constraints (denoted by \textcolor{red}{red} lines)}
\label{fig:sopflow}
\end{figure}

The formulation for the single period stochastic optimal power flow is given in \ref{eq:sopflow_start}--\ref{eq:sopflow_end}. This formulation is similar to single-period SCOPFLOW. The difference is \sopflow objective is weighted by $\pi_s$ for each scenario.

\begin{align}
\centering
\text{min}&\sum_{s \in \mathcal{S}}\pi_s f(x_s)  \label{eq:sopflow_start}\\
&\text{s.t.} \nonumber \\
&~g(x_s) = 0,                              \\
&~h(x_s) \le 0,                            \\
x^- & \le x_s \le x^+,                     \\
-\delta_s{x} & \le x_s - x_0 \le \delta_s{x},~ s\neq 0
\label{eq:sopflow_end}
\end{align}

here, ${S}$ is the set of scenarios with the base-scenario denoted by subscript 0. While there is no real notion of a "base" scenario, our current implementation assumes it as the scenario with the most weight or probability. The coupling constraints between the base-case and scenario subproblem are ramping constraints for the generators.

The single period, no contingency \sopflow can be solved on a single processor (serial) with \ipopt, in parallel using \hiop~ primal decomposition approach, or in an embarassingingly parallel way with the EMPAR solver.

\subsection{Single period, multiple contingencies}
The aim of the single period, contingency-constrained stochastic OPF is to optimize the grid dispatch to ensure the grid is secure for all the wind forecast deviations and for all the contingencies considered therein. There are two variations of this problem implemented in \exago differing in how the contingency-scenario pairs are set up. 

\subsubsection{Full stochastic contingency-constrained structure}
(\ref{eq:sopflow_full_start})-(\ref{eq:sopflow_full_end}) describes the formulation for the full stochastic contingency-constrained OPF formulation and its structure is illustrated in Figure \ref{fig:sopflow_ctgc1} with a two scenarios- two contingency example..

\definecolor{lavander}{cmyk}{0,0.48,0,0}
\definecolor{violet}{cmyk}{0.79,0.88,0,0}
\definecolor{burntorange}{cmyk}{0,0.52,1,0}

\def\lav{lavander!90}
\def\oran{orange!30}

\tikzstyle{contingency}=[draw,circle,violet,bottom color=\lav,
                  top color= white, text=violet,minimum width=20pt]
\tikzstyle{base}=[draw,circle,burntorange, left color=\oran,
                       text=violet,minimum width=20pt]
                       
\tikzstyle{cedge}=[color=red]

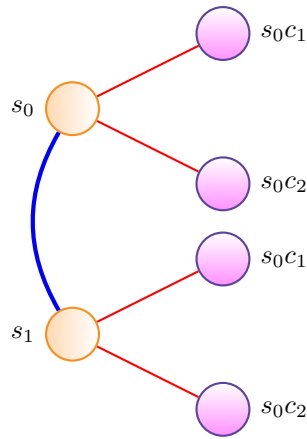
\begin{figure}[h!]
\centering
\begin{tikzpicture}[auto, thick]
  % Place base case
  \node[base,label=left:$s_0$] (s0c0) at (0,0) {};
  
  \node[contingency,label=right:$s_0c_1$] (s0c1) at (2,1) {};
  \node[contingency,label=right:$s_0c_2$] (s0c2) at (2,-1) {};
  
  \node[base,label=left:$s_1$] (s1c0) at (0,-3) {};
  
  \node[contingency,label=right:$s_0c_1$] (s1c1) at (2,-2) {};
  \node[contingency,label=right:$s_0c_2$] (s1c2) at (2,-4) {};
  
  \path[cedge] (s0c0) edge (s0c1);
  \path[cedge] (s0c0) edge (s0c2);
  \path[cedge] (s1c0) edge (s1c1);
  \path[cedge] (s1c0) edge (s1c2);
  
  \path (s0c0) edge [bend right,ultra thick,color=blue] (s1c0);

\end{tikzpicture}
\caption{Full stochastic contingency-constrained optimal power flow structure}
\label{fig:sopflow_ctgc1}
\end{figure}

$s_0$ and $s_1$ represent the base case for the two scenarios with two contingencies $c_1$ and $c_2$. Each of the contingency states is coupled with its scenario base-case through ramping constraints (denoted by \textcolor{red}{red} lines) and each scenario is coupled with each other through having a common real power set-point for generators \textcolor{blue}{blue} (see \cite{exago_manual} for the generator set-point formulation). This structure results in a three-stage optimization problem that is difficult to solve in parallel since most of the optimization solvers cater to two-stage optimization problems only. Hence, \sopflow only supports solving this problem on a single processor using \ipopt currently.

\begin{align}
\centering
\min &\sum_{s \in \mathcal{S}}\pi_s \sum_{c \in \mathcal{C}}f(x_{s,c})  \label{eq:sopflow_full_start}\\
&\text{s.t.} \nonumber \\
&~g(x_{s,c}) = 0,                              \\
&~h(x_{s,c}) \le 0,                            \\
x^- & \le x_{s,c} \le x^+,                     \\
-\delta_s{x} & \le x_{s,0} - x_{0,0} \le \delta_s{x},~ s\neq 0 \\
-\delta_c{x} & \le x_{s,c} - x_{s,0} \le \delta_c{x},~ c\neq 0
\label{eq:sopflow_full_end}
\end{align}

\subsection{Flattened stochastic contingency-constrained structure}
This variation is a relaxation of the full structure obtained by flattening the contingency-scenario pairs to reduce the problem to a two-stage optimization problem. The formulation is given in (\ref{eq:sopflow_full_start})-(\ref{eq:sopflow_full_end}) and its structure is illustrated in Figure \ref{fig:sopflow_ctgc2}. 

\definecolor{lavander}{cmyk}{0,0.48,0,0}
\definecolor{violet}{cmyk}{0.79,0.88,0,0}
\definecolor{burntorange}{cmyk}{0,0.52,1,0}

\def\lav{lavander!90}
\def\oran{orange!30}

\tikzstyle{contingency}=[draw,circle,violet,bottom color=\lav,
                  top color= white, text=violet,minimum width=20pt]
\tikzstyle{base}=[draw,circle,burntorange, left color=\oran,
                       text=violet,minimum width=20pt]
                       
\tikzstyle{cedge}=[color=red]

\begin{figure}[h!]
\centering
\begin{tikzpicture}[auto, thick]
  % Place base case
  \node[base,label=left:$s_0$] (s0c0) at (0,-2) {};
  
  \node[contingency,label=right:$s_0c_1$] (s0c1) at (2,0) {};
  \node[contingency,label=right:$s_0c_2$] (s0c2) at (2,-1) {};
  
  \node[contingency,label=right:$s_1$] (s1c0) at (2,-2) {};
  
  \node[contingency,label=right:$s_1c_1$] (s1c1) at (2,-3) {};
  \node[contingency,label=right:$s_1c_2$] (s1c2) at (2,-4) {};
  
  \path[cedge] (s0c0) edge (s0c1);
  \path[cedge] (s0c0) edge (s0c2);
  \path[cedge] (s0c0) edge (s1c1);
  \path[cedge] (s0c0) edge (s1c2);
  
  \path[cedge] (s0c0) edge (s1c0);

\end{tikzpicture}
\caption{Flattened stochastic contingency-constrained optimal power flow structure}
\label{fig:sopflow_ctgc2}
\end{figure}
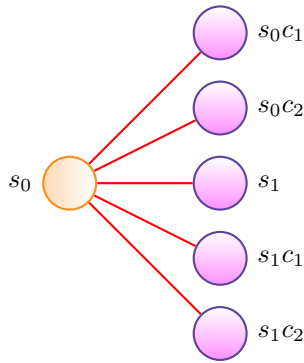

\begin{figure}[h!]
\centering
  {\includegraphics[width=\columnwidth,trim = 0cm 0cm 0cm 13cm,clip]{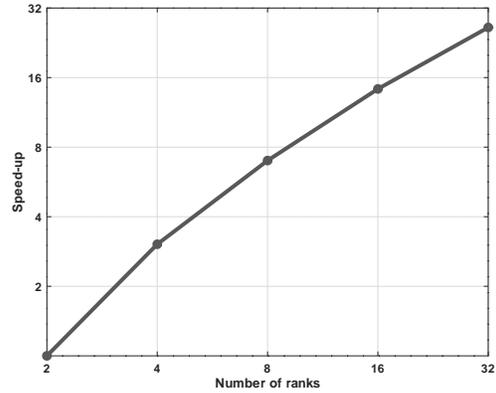}}
  \caption{Scalability of SOPFLOW using primal decomposition approach. The HIOP solver was used for solving SOPFLOW on the ACTIVSg200 system with 5 scenarios and 100 contingencies.}
  \label{fig:sopflow_hiop_scaling}
\end{figure}

The scenarios and the contingencies are flattened to obtain a standard two-stage optimization structure. This is done by choosing the most probable scenario as the base-case and coupling (denoted by \textcolor{red}{red}) it with all the other scenario-contingency pairs. This problem variation can be solved both in serial (using \ipopt or \hiop) or in parallel (using \hiop). The difference between the full structure and the flattened structure can be seen in the last constraint in their respective formulations. While the full structture has coupling constraints for contingencies with their respective "base" scenario, the flattened structured has all contingencies (and scenarios) coupled to a single base scenario. Figure \ref{fig:sopflow_hiop_scaling} shows the strong-scaling of SOPFLOW with the flattened contingency-structure on the ACTIVSg200 bus system with 5 wind scenarios each with 100 contingencies.

\begin{align}
\centering
\text{min}&\sum_{s \in \mathcal{S}}\pi_s \sum_{c \in \mathcal{C}}f(x_{s,c})  \label{eq:sopflow_flat_start}\\
&\text{s.t.} \nonumber \\
&~g(x_{s,c}) = 0,                              \\
&~h(x_{s,c}) \le 0,                            \\
x^- & \le x_{s,c} \le x^+,                     \\
-\delta_s{x} & \le x_{s,0} - x_{0,0} \le \delta_s{x},~ s\neq 0 \\
-\delta_c{x} & \le x_{s,c} - x_{0,0} \le \delta_c{x},~ c\neq 0
\label{eq:sopflow_flat_end}
\end{align}

\subsection{Multiperiod, multi-contingency}
This is the most detailed formulation involving stochasticity, security constraints, and time, i.e. all the three dimensions of uncertainty. The full formulation for the stochastic security-constrained multi-period optimal power flow is given in (\ref{eq:sctopflow_start}) -- (\ref{eq:sctopflow_end}). In this formulation, the objective is to reduce the expected cost, where $f(x_{s,c,t})$ is the cost for scenario $s$ with contingency $c$ at time $t$. $\pi_s$ is the probability of scenario $s$.

\begin{align}
\centering
\text{min}&~\sum_{s \in \mathcal{S}}\pi_s\sum_{c \in \mathcal{C}}\sum_{t \in \mathcal{T}}f(x_{s,c,t})  \label{eq:sctopflow_start}\\
&\text{s.t.} \nonumber \\
&~g(x_{s,c,t}) = 0                                         \\
&~h(x_{s,c,t}) \le 0                                   \\
x^- & \le x_{s,c,t} \le x^+                                \\
-\delta_t{x} & \le x_{s,c,t} - x_{s,c,t-\Delta{t}} \le \delta_t{x},~t \neq 0 \label{eq:sctopflow_time_coupling}\\
-\delta_c{x} & \le x_{s,c,0} - x_{s,0,0} \le \delta_c{x},~c \neq 0
\label{eq:sctopflow_contingency_coupling} \\
-\delta_s{x} & \le x_{s,0,0} - x_{0,0,0} \le \delta_s{x},~s \neq 0
\label{eq:sctopflow_end}
\end{align}
here, $c \in \mathcal{C}$, $t \in \mathcal{T}$, and $s \in \mathcal{S}$ are the indices for contingencies, time-step, and scenarios, respectively.

An illustration of \sopflow is shown in Fig. \ref{fig:sctopflow} for a case with two scenarios $s_0$ and $s_1$. Each scenario has two contingencies $c_0$, $c_1$, and each contingency has four time-periods.

\definecolor{lavander}{cmyk}{0,0.48,0,0}
\definecolor{violet}{cmyk}{0.79,0.88,0,0}
\definecolor{burntorange}{cmyk}{0,0.52,1,0}

\def\lav{lavander!90}
\def\oran{orange!30}

\tikzstyle{time}=[draw,circle,violet,bottom color=\lav,
                  top color= white, text=violet,minimum width=20pt]
                  
\tikzstyle{base}=[draw,circle,burntorange, left color=\oran,
                       text=violet,minimum width=20pt]

\begin{figure}[h!]
\centering
\scalebox{0.9}{
\begin{tikzpicture}[auto, thick]
  % scenario 0
  \node[base,label=below:$t_0$] (s0c0t0) at (0,2) {};
  \node[time,label=below:$t_1$] (s0c0t1) at (2,2) {};
  \node[time,label=below:$t_2$] (s0c0t2) at (4,2) {};
  \node[time,label=below:$t_3$] (s0c0t3) at (6,2) {};
  
  \path (s0c0t0) edge (s0c0t1);
  \path (s0c0t1) edge (s0c0t2);
  \path (s0c0t2) edge (s0c0t3);
  
  % contingency rectangle from (-0.5,1.5) to (6.5,2.5)
   \node[draw,rectangle,dashed,minimum width=7cm,minimum height=1cm,label=left:{$\mathbf{c_0}$}] (s0) at (3.0,2.0) {};
  
  \node[time,label=below:$t_0$] (s0c1t0) at (0,0) {};
  \node[time,label=below:$t_1$] (s0c1t1) at (2,0) {};
  \node[time,label=below:$t_2$] (s0c1t2) at (4,0) {};
  \node[time,label=below:$t_3$] (s0c1t3) at (6,0) {};
  
  \path (s0c1t0) edge (s0c1t1);
  \path (s0c1t1) edge (s0c1t2);
  \path (s0c1t2) edge (s0c1t3);
  
  % contingency rectangle from (-0.5,-0.5) to (6.5,0.5)
   \node[draw,rectangle,dashed,minimum width=7cm,minimum height=1cm,label=left:{$\mathbf{c_1}$}] (s0) at (3.0,0.0) {};
  
   \path (s0c0t0) edge [bend right,color=red] (s0c1t0);
  
  % scenario rectangle from (-1.0,-0.75) to (7.0,2.75)
   \node[draw,rectangle,dotted,minimum width=8cm,minimum height=3.5cm,label=left:{$\mathbf{s_0}$}] (s0) at (3.0,1.0) {}; 
  
  % scenario 1
  \node[base,label=below:$t_0$] (s1c0t0) at (0,4) {};
  \node[time,label=below:$t_1$] (s1c0t1) at (2,4) {};
  \node[time,label=below:$t_2$] (s1c0t2) at (4,4) {};
  \node[time,label=below:$t_3$] (s1c0t3) at (6,4) {};
  
  \path (s1c0t0) edge (s1c0t1);
  \path (s1c0t1) edge (s1c0t2);
  \path (s1c0t2) edge (s1c0t3);
  
   % contingency rectangle from (-0.5,3.5) to (6.5,4.5)
   \node[draw,rectangle,dashed,minimum width=7cm,minimum height=1cm,label=left:{$\mathbf{c_0}$}] (s0) at (3.0,4.0) {};
  
  \node[time,label=below:$t_0$] (s1c1t0) at (0,6) {};
  \node[time,label=below:$t_1$] (s1c1t1) at (2,6) {};
  \node[time,label=below:$t_2$] (s1c1t2) at (4,6) {};
  \node[time,label=below:$t_3$] (s1c1t3) at (6,6) {};
  
  \path (s1c1t0) edge (s1c1t1);
  \path (s1c1t1) edge (s1c1t2);
  \path (s1c1t2) edge (s1c1t3);
  
   % contingency rectangle from (-0.5,5.5) to (6.5,6.5)
   \node[draw,rectangle,dashed,minimum width=7cm,minimum height=1cm,label=left:{$\mathbf{c_1}$}] (s0) at (3.0,6.0) {};
  
  \path (s1c0t0) edge [bend left,color=red] (s1c1t0);
  
  \path (s0c0t0) edge [bend left,ultra thick,color=blue] (s1c0t0);
  
   % scenario rectangle from (-1.0,3.75) to (7.0,6.75)
   \node[draw,rectangle,dotted,minimum width=8cm,minimum height=3.5cm,label=left:{$\mathbf{s_1}$}] (s1) at (3.0,5.0) {};

\end{tikzpicture}
}
\caption{Stochastic multi-period contingency constrained example with two scenarios $s_0$ and $s_1$. Each scenario has two contingencies $c_0$ and $c_1$ and each contingency consists of four time-periods $t_0$, $t_1$, $t_2$, $t_3$.  State $s_0,c_0,t_0$ represents the base case (no contingency) case for the two scenarios.  We assume that any contingency is incident at the first time-step, i.e., at $t_0$. Thus, the contingency states $c_1,t_0$ is coupled with the no-contingency state $c_0,t_0$ at time $t_0$ for both the scenarios. The {\textcolor{red}{red}} line denotes the coupling between the contingency and the no-contingency states.The {\textcolor{blue}{blue}} line denotes the coupling between the scenarios}
\label{fig:sctopflow}
\end{figure}
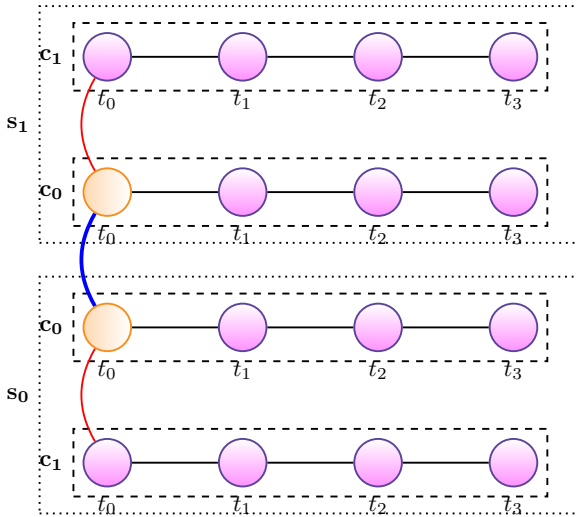

\sopflow currently only supports \ipopt solver to solve the stochastic security-constrained multi-period optimal power flow problem.

\section{Conclusions and Future Work}
Dealing with increased uncertainties will be important to address reliability issues in a deep decarbonized grid with low-inertia and extensive penetration off renewable energy sources. The ExaGO toolkit aims to solve large-scale stochastic, security-constrained, multi-period ACOPF problems on high-performance computers. It has high-fidelity models with scalable numerical algorithms. Moreover, it has a well-designed API through which new models and solvers can be interfaced. ExaGO has been validated against PowerWorld \cite{powerworld} on several test networks and we are continually validating it on more networks and applications. It is in active development and several new features are to be added to the library including new modeling capabilities (storage, flexible load), improving robustness for large-scale optimization problems,and high-performance implementation of multi-period, multi-contingency stochastic ACOPF, and expanding its solver capabilities. 

\section*{Acknowledgements}
This research was supported by the Exascale Computing Project (ECP), Project Number: 17-SC-20-SC, a collaborative effort of two DOE organizations—the Office of Science and the National Nuclear Security Administration—responsible for the planning and preparation of a capable exascale ecosystem—including software, applications, hardware, advanced system engineering, and early testbed platforms—to support the nation's exascale computing imperative.

The authors thank Cosmin Petra, Naiyuan Chiang, and Jingyi Wang for their guidance and support on the usage of \hiop~ 

The authors also want to thank Phil Roth, Christopher Oehmen, and Lori Ross O'Neil for their support on this work.

\bibliographystyle{IEEEtran}
\bibliography{bib1,bib2,bib_slaven}

\vfill
\end{document}